\documentclass{ws-mpla}

\usepackage[superscript]{cite}

\newcommand{\Tr}{\mathrm{Tr}}

\newcommand{\bea}{\begin{eqnarray}}
\newcommand{\eea}{\end{eqnarray}}

\newcommand{\hspp}{\hspace{-.1cm}}
\newcommand{\nn}{\nonumber \\}

\def\slash#1{\setbox0=\hbox{$#1$}  
   \dimen0=\wd0     
   \setbox1=\hbox{/} \dimen1=\wd1  
   \ifdim\dimen0>\dimen1   
      \rlap{\hbox to \dimen0{\hfil/\hfil}} 
      #1     
   \else     
      \rlap{\hbox to \dimen1{\hfil$#1$\hfil}} 
      /      
   \fi}      %

\makeatother
\begin{document} 
 
\markboth{Leonard Gamberg  and Marc Schlegel} 
{Final state interactions and the transverse 
structure of the pion} 
 
\catchline{}{}{}{}{} 
 
\title{FINAL STATE INTERACTIONS AND THE TRANSVERSE STRUCTURE
OF THE PION} 

\author{LEONARD GAMBERG$^{1,2}$ and MARC SCHLEGEL$^3$}

 \address{
  $^1$ Institute for Nuclear Theory, University of Washington 
\\ Seattle,  Washington 98195-1550, USA
\\
  $^2$Division of Science, Penn State University-Berks \\ 
Reading, Pennsylvania 19083, USA
\\
  $^3$ 
       Thomas Jefferson National Accelerator Facility, \\ Newport News, VA 23606, USA}
 
\maketitle 
 
\pub{Received (28 October 2009)}{}
 
\begin{abstract} 
In the factorized picture of semi-inclusive deep inelastic scattering
the naive  time reversal-odd parton distributions exist by virtue
of the gauge link which is intrinsic to their definition.
 The link structure describes
initial/final-state interactions of the active parton due to soft
gluon exchanges with the target remnant. Though these  interactions
are non-perturbative,   calculations of
 final-state interaction have been performed in a
 perturbative one-gluon approximation. We 
 include higher-order
 contributions 
by applying non-perturbative eikonal methods to  calculate  
the Boer-Mulders function of the pion.
Using this
framework  we explore 
under what conditions the Boer Mulders function  can be described in
terms of factorization of final state interactions and a spatial distortion.

\end{abstract} 

\ccode{13.88.+e, 
      13.85.Ni,  
      13.60.-r,  
      13.85.Qk}  
 
\section{T-odd PDFs, Gluonic Poles and The Lensing Function}
\label{intro}
Over the past two decades the transverse partonic structure of hadrons
has been the subject of a great deal of theoretical and experimental
study. Central to these investigations are the early observations
of large transverse single spin asymmetries (TSSAs) in 
inclusive hadron production from proton-proton scattering over a wide
range of beam energies~\cite{Bunce:1976yb,Dragoset:1978gg,Antille:1980th,Adams:1991cs}. Recently TSSAs have been observed in lepton-hadron semi-inclusive
deep inelastic scattering 
(SIDIS)\cite{Airapetian:1999tv,Airapetian:2004tw,Avakian:2003pk,Avakian:2005ps,Schill:2008ra} as well
as in inclusive production of pseudo-scalar mesons from in proton-proton collisions\cite{Adams:2003fx,Adler:2005in,Arsene:2008mi,Abelev:2008qb}. While the naive parton model predicts
that transverse polarization effects are 
trvial in the helicity limit\cite{Kane:1978nd}, 
Efremov and Teryaev demonstrated\cite{Efremov:1984ip,Efremov:1981sh} 
that soft gluonic and fermionic poles contribute to
multiparton correlation functions  resulting in non-trivial twist-three
transverse polarization effects in 
this limit\cite{Qiu:1991pp,Qiu:1991wg}. In addition
theoretical work on transversity\cite{Ralston:1979ys,Jaffe:1991kp,Collins:1992kk}
indicates that transverse polarization effects can appear at leading
twist. Two explanations to account for TSSAs in QCD have emerged which
are based on the twist-three\cite{Qiu:1991pp,Qiu:1991wg} and 
twist-two\cite{Sivers:1989cc,Collins:1992kk,Anselmino:1994tv,Mulders:1995dh,Boer:1997nt}
approaches. Recently, a coherent picture has emerged which describes
TSSAs in a kinematic regime where the 
two approaches are expected to have a common description\cite{Boer:2003cm,Ji:2006ub,Ji:2006br,Bacchetta:2008xw}.

In the factorized picture of 
semi-inclusive deep inelastic scattering\cite{Mulders:1995dh,Ji:2004wu}
at small transverse momenta 
$P_T\sim k_T << \sqrt{Q^2}$
the Sivers effect describes a 
transverse target spin-$S_T$ asymmetry 
through the ``naive'' T-odd structure,
$\Delta f(x,\vec{k}_{T})\sim S_{T}\cdot(P\times\vec{k}_{T})f_{1T}^{\perp}(x,k_{T}^{2})$\cite{Sivers:1989cc,Sivers:1990fh}.  
For an unpolarized target with transversely polarized quarks-$s_{T}$,
the Boer-Mulders function\cite{Boer:1997nt} is $\Delta h(x,\vec{k}_{T})\sim s_{T}\cdot(P\times\vec{k}_{T})h_{1}^{\perp}(x,k_{T}^{2})$.  Many 
studies have been performed to model the T-odd PDFs in terms of the
FSIs where 
soft gluon rescattering effects 
are approximated 
by perturbative 
one-gluon 
exchange\cite{Brodsky:2002cx,Ji:2002aa,Goldstein:2002vv,Boer:2002ju,Gamberg:2003ey,Gamberg:2003eg,Bacchetta:2003rz,Lu:2004hu,Gamberg:2007wm,Bacchetta:2008af}.
We improve  this approximation by 
applying  non-perturbative eikonal methods to 
calculate higher-order gluonic contributions 
from the gauge link 
 in the spectator framework~\cite{Gamberg:2009uk}. 
In the context of these 
higher order contributions 
we  perform a quantitative study of approximate relations
between TMDs and GPDs. In particular, we 
explore under what conditions the 
T-odd PDFs can be described via factorization of FSI and spatial
distortion of impact parameter space PDFs\cite{Burkardt:2002hr}.
While such relations
are fulfilled from lowest order contributions in  field-theoretical
spectator models\cite{Burkardt:2003je,Meissner:2007rx} a model-independent
analysis of 
generalized parton correlation functions (GPCFs)\cite{Belitsky:2003nz}
indicates that the Sivers function and the helicity flip GPD $E$
are projected from independent GPCFs. A similar result holds for the
Boer-Mulders function for a spin zero target\cite{Meissner:2008ay}.
 From phenomenology, however it essentially unknown  whether the
proposed factorization 
 is a good approximation. Here we focus on the transverse structure
of the pion through the impact parameter GPD and  
the Boer Mulders function for which very little know.
Recent lattice calculations indicate that the spatial asymmetry of
transversely polarized quarks in the pion is quite similar in magnitude
to that of quarks in the nucleon\cite{Brommel:2007xd}.

The field-theoretical definition of transverse-momentum dependent
(TMD) parton distributions in terms of hadronic matrix elements of
quark operators 
for  spin-1/2 hadron with momentum $P$ and spin $S$ was
presented in Refs.\cite{Mulders:1995dh,Goeke:2005hb,Bacchetta:2008xw}. It is
straightforward
 to obtain the TMDs for a spin-0 hadron from that. One encounters
two leading twist TMDs for a pion, the distribution for unpolarized quarks $f_1$,
and the distribution of transversely polarized quarks $h_1^{\perp}$, the 
Boer-Mulders function.
Adopting the infinite-momentum frame where the hadron moves relativistically
along the positive $z$-axis such that the target momentum $P$ has
a large plus component $P^{+}$ and no transverse component 
the Boer-Mulders function, defined in SIDIS is 
\begin{equation}
\frac{\epsilon_T^{ij}k_T^j h_1^{\perp}(x,\vec{k}_{T}^2)}{m_\pi}
\hspp=\hspp\int\frac{dz^{-}d^{2}z_{T}}{4(2\pi)^{3}}
\mathrm{e}^{{\scriptstyle ixP^{+}z^{-}-i\vec{k}_{T}\cdot\vec{z}}}
\langle P|\bar{q}_{j}(0) [{\scriptstyle 0 ; \infty n}] i\sigma^{i+}\gamma_5[{\scriptstyle \infty n + z_T ; z}]\, q_{i}(z) |P\rangle. \label{eq:Correlator}
\nn
\end{equation}
The light-like vector $n$ represents a specific direction on
the light-cone $n^{\mu}=(1,0,0)$
where we define
the light cone components of a 4-vector $a^{\pm}=1/\sqrt{2}(a^{0}\pm a^{3})$,
$a^\mu=(a^-,a^+,a^\perp)$.  $[x\,;\, y]$ denotes a gauge link operator 
connecting the two locations $x$ and $y$.  
We work in a covariant gauge where a transverse
gauge link at light-cone infinity is negligible. The gauge link in
(\ref{eq:Correlator}) is interpreted physically as FSIs of the active
quark with the target remnants\cite{Brodsky:2002cx,Collins:2002kn}
and is necessary for {}``naive'' time-reversal odd (T-odd) TMDs\cite{Sivers:1989cc,Sivers:1990fh,Boer:1997nt} 
to exist\cite{Collins:2002kn}.
The Boer-Mulders function appears in the factorized description of
semi-inclusive processes such as SIDIS\cite{Mulders:1995dh,Boer:1997nt,Boer:2003cm,Bacchetta:2006tn,Ji:2004wu,Ji:2004xq,Collins:2004nx,Tangerman:1994bb,Boer:1999mm,Arnold:2008kf}
in terms of the first $k_{T}$-moment, 
$2m_{\pi}^{2}h_{1}^{\perp(1)}(x)=\int d^{2}k_{T}\,\vec{k}_{T}^{2}\, h_{1}^{\perp}(x,\vec{k}_{T}^{2})$.
Transforming the two pion states in Eq.~(\ref{eq:Correlator}) into a mixed coordinate-momentum representation specified
by the impact parameter $b_T$\cite{Burkardt:2003uw,Meissner:2007rx}
results in an impact parameter representation
for the gluonic pole matrix element\cite{Boer:2003cm}
$\langle k_{T}\rangle(x) =m_{\pi}h_{1}^{\perp(1)}(x)$,
\begin{eqnarray} 
\langle k_{T}\rangle(x)=\int d^2b_T\frac{dz^{-}}{4(2\pi)}\mathrm{e}^{ixP^{+}z^{-}}
\langle P^+,\vec{0}_T|\,\bar{q}(z_1)\,[z_1\,;\,z_2] I^{i}(z_2) \sigma^{i+} q(z_2)\,|P^+,\vec{0}_T\rangle. \label{eq:RelMI}
\end{eqnarray}
Here, the impact parameter $b_{T}$ is hidden in the arguments of
the quark fields, $z_{1/2}^\mu=\mp\frac{z^{-}}{2}n^\mu+b_T^\mu$
and    $b_T^\mu=(0,b_T^1,b_T^2,0)$.
 The operator
$I^{i}$ originates from the time-reversal behavior of the FSIs
 written in terms of the  gauge link in (\ref{eq:Correlator}) and
the  field strength tensor $F^{\mu\nu}$,
\begin{equation}
2I^{i}(z_2)=\int dy^{-}\,[z_2\,;\, y]\, gF^{+i}(y)\,[y\,;\,z_2],
\end{equation}
with $y^\mu=y^- n^\mu+b_T^\mu$.

Turning our attention to GPDs of a pion, they 
are represented by an off-diagonal matrix element of a
quark-quark operator defined on the light-cone\cite{Diehl:2003ny,Goeke:2001tz,Belitsky:2005qn}. 
One encounters two leading twist GPDs for a pion, a chirally-even GPD $F_1^\pi$ and
the chiral odd 
 GPDs $H_{1}^{\pi}$\cite{Meissner:2008ay}.  
We use the symmetric conventions for the kinematics for GPDs\cite{Diehl:2003ny},
$P=\frac{1}{2}(p+p^{\prime})$ and $\Delta=p^{\prime}-p$ where 
$\Delta^{+}=-2\xi P^{+}$, and $t=\Delta^{2}$.
The impact parameter GPDs are obtained from the ordinary GPDs via
a Fourier-transform of the transverse momentum transfer $\vec{\Delta}_{T}$
at zero skewness $\xi=0$. The chirally-odd impact parameter GPD $\mathcal{H}_{1}^{\pi}$
 is expressed as
\begin{eqnarray}
\int\frac{dz^{-}}{2(2\pi)}\mathrm{e}^{ixP^{+}z^{-}}\langle P^{+},\vec{0}_{T}|\,\bar{q}(z_{1})[z_{1};z_{2}]\sigma^{+i}q(z_{2})\,|P^{+},\vec{0}_{T}\rangle
=\frac{2 b_{T}^{i}}{m_{\pi}}\,\frac{\partial}{\partial\vec{b}_{T}^{2}}\mathcal{H}_{1}^{\pi}(x,\vec{b}_{T}^{2}).
\end{eqnarray}
$\mathcal{H}_{1}^{\pi}$ describes how transversely polarized
quarks are distributed in a plane transverse
to the direction of motion. 
This distribution represents
transverse space distortion due 
to spin-orbit correlations\cite{Burkardt:2005hp,Diehl:2005jf,Brommel:2007xd}.
A comparison of  the
first moment of the Boer Mulders function (\ref{eq:RelMI}) and the
impact parameter GPD $\mathcal{H}_{1}^{\pi}$
reveals that they differ 
by  the operator $I^{i}$ which represents the FSIs. 
In various model calculations, 
the FSIs 
are approximated such that\cite{Burkardt:2003je,Burkardt:2003uw,Lu:2006kt,Meissner:2007rx}
the two effects of a distortion of the transverse space
parton distribution and the FSIs factorize
resulting in the  quantitative relation
\begin{equation}
m_{\pi}\epsilon_{T}^{ij}h_{1}^{\perp(1)}(x)\simeq\int d^{2}b_{T}\,\mathcal{I}^{j}(x,\vec{b}_{T})\frac{\epsilon_{T}^{il}b_{T}^{l}}{m_{\pi}}\,\frac{\partial}{\partial\vec{b}_{T}^{2}}\mathcal{H}_{1}^{\pi}(x,\vec{b}_{T}^{2}),\label{eq:Relation}
\end{equation}
where $\mathcal{I}$ is the so-called ``quantum chromodynamic 
lensing function''\cite{Burkardt:2003uw}. This factorization  doesn't hold
in general\cite{Meissner:2008ay,Meissner:2009ww}. For example, this relation breaks down when
 the quark fields and the operator $I$ in (\ref{eq:RelMI}) {}``interact''
via quantum fluctuations (because they are interacting Heisenberg
operators). However, it  unknown if (\ref{eq:Relation})
is a good phenomenological approximation.  We estimate 
the size of the lensing function using
non-perturbative eikonal methods\cite{Abarbanel:1969ek,Fried:2000hj}
to calculate higher-order 
 gluon contributions from the gauge
link.  Up till now the relation (\ref{eq:Relation}) was used
to predict the sign of T-odd TMDs in conjunction with numbers
for the $u$- and $d$-quark contributions to the anomalous magnetic
moment of the nucleon and the assumption that final state interactions
are attractive\cite{Burkardt:2005hp}. 
\section{TMD - GPD Relation for a Pion the  Spectator Framework}
We  focus our attention in the following on a pion in a valence
quark-type configuration that one would 
expect for relatively large Bjorken $x$. 
Thus, we only take valence quark wave functions
into account and neglect higher Fock states. 
Assuming an antiquark spectator one can express the pion Boer-Mulders function~(\ref{eq:Correlator}) as 
\begin{equation}  
\epsilon_T^{ij} k_T^j h_1^{\perp}(x,\vec{k}_{T}^2)  = \frac{m_\pi}{8(2\pi)^3(1-x)P^+}\sum_{\sigma,d} \bar{W}i\sigma^{i+}\gamma_5 W,\label{eq:PhiSpectator}
\end{equation} 
where $ W_i^{\alpha, \delta}(P,k;\sigma)=\langle P-k,\sigma,\delta|\,[\infty n\,;\,0]^{\alpha \beta}\, q_{i}^{\beta}(0)\,|P\rangle$ and 
$\sigma$ and $\delta$ represent the helicity and color of the intermediate spectator antiquark.
\begin{figure}
\begin{center}
\includegraphics[scale=0.7]{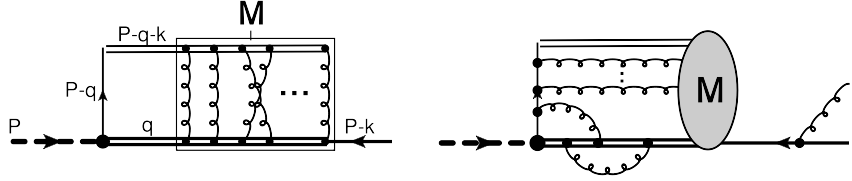}
\caption{\scriptsize The amplitude $W$ including FSIs between re-scattered
eikonalized quark and antiquark. The FSIs
are described by a non-perturbative scattering amplitude $M$ that is 
calculated in a generalized ladder approximation.
Gluon interactions as shown in the second  diagram are not taken into account (see text).\label{fig:TMD-amplitude-including}}
\end{center}
\end{figure}
We model the matrix element for $W$ 
by a diagram shown in Fig. \ref{fig:TMD-amplitude-including}. 
where 
the final state interactions \--- generated by the gauge link in $W$
\--- are described by a non-perturbative amputated scattering amplitude
$({M})_{\gamma \delta}^{\alpha\beta}$ with $\beta,\,\alpha$
($\gamma,\,\delta$) color indices of incoming and outgoing quark
(anti-quark), respectively.  We neglect classes of gluon exchanges in the second diagram in Fig. \ref{fig:TMD-amplitude-including}
represented by the gluon rungs
since they would be attributed to the ``interaction'' between the
quark fields and the operator $I$ in (\ref{eq:RelMI}). They
lead to terms which break the relation (\ref{eq:Relation}). We also
neglect 
real gluon emission and (self)-interactions
of quark and antiquark lines the second diagram in Fig. \ref{fig:TMD-amplitude-including} 
since they represent radiative corrections of the GPD and are effectively modeled in
terms of spectator masses and phenomenological vertex functions.  
The pion-quark vertex is governed by the interaction 
Lagrangian
$\mathcal{L}=-g_{\pi}/\sqrt{N_c} \delta^{\alpha \beta}\bar{q}^{\alpha}\gamma_{5}\vec{\tau}\cdot\vec{\varphi}q^{\beta}$
and  we allow the coupling constant $g_{\pi}$
to depend on the momentum of the active quark in order to account
for the compositeness of the hadron and to suppress large quark virtualities.
Phenomenological vertex functions in connection with spectator
models have been used frequently in the literature\cite{Jakob:1997wg,Gamberg:2007wm,Bacchetta:2008af}. Applying
the  Feynman rules it is then possible to find an expression
for the matrix element $W$  from the first diagram in Fig.
\ref{fig:TMD-amplitude-including}. We find 
\begin{eqnarray}
\hspace{-0.5cm}W_{i,\sigma}^{\alpha\beta}(P,k) &=&\frac{-i\tau}{\sqrt{N_c}}\Bigg[\delta^{\alpha\beta}g_{\pi}(k^{2})\frac{\left[(\slash k+m_{q})v(P_s,\sigma)\right]_{i}}{k^{2}-m_{q}^{2}+i0}-\int\frac{d^{4}q}{(2\pi)^{4}}\nn
 && \hspace{-1.75cm}\frac{ g_{\pi}\left((P-q)^{2}\right)
\left[(\slash P-\slash q+m_{q})\gamma_{5}(-\slash q+m_{s})
\left({M}\right)_{\delta\beta}^{\alpha\delta}(q,P_s)v(P_s,\sigma)\right]_{i}}{\left[n\cdot(P_s-q)+i0\right]\left[(P-q)^{2}-m_{q}^{2}+i0\right]\left[q^{2}-m_{s}^{2}+i0\right]}\Bigg],
\label{eq:StartW}
\end{eqnarray}
where $P_s\equiv P-k$ is the spectator momentum.
The first term in (\ref{eq:StartW}) represents 
the  contribution without
final state interactions while the second term corresponds
to the first diagram in Fig. \ref{fig:TMD-amplitude-including}.
We then express the FSIs through the amputated quark - anti-quark scattering
amplitude ${M}$.  Here \emph{both} incoming quark and anti-quark
are subject to the 
eikonal approximation (see, e.g. Ref.~\refcite{Fried:1990ei} and references therein).
While the active quark undergoes a natural eikonalization
for a massless fermion since it represents the gauge link contribution,
the eikonalization for a massive spectator fermion is a simplification
that can  be justified by the physical picture of partons in an
infinite momentum frame. The eikonalization of a massive
fermion can be traced back to the Nordsieck-Bloch approximation\cite{Bloch:1937pw}  which describes a highly energetic
helicity conserving fermion undergoing multiple scattering
undergoing very small momentum transfer. 
In this approximation 
the Dirac vertex 
structure, $\bar{u}(p_1)\gamma^\mu u(p_2)\sim p_\mu/m$ where
$(p_1+p_2)/2\equiv p$. For a massive anti-fermion
one identifies the velocity $v^\mu=-p^{\mu}/m$, 
and the numerator of  a fermion propagator becomes  $i(-\slash p+m)\rightarrow i(-\bar{n}\cdot p+m)$.

We proceed by performing a contour-integration of the light-cone
 loop-momentum, $q^{-}$  in Eq.(\ref{eq:StartW}). In doing
so we only consider poles which originate from the 
denominators in (\ref{eq:StartW}).
This assumes that the scattering amplitude ${M}$ does not
contain poles in $q^{-}$. 
This assumption is not necessarily true even for a one gluon 
exchange approximation.
As shown in Refs.~\refcite{Gamberg:2006ru,Gamberg:2007wm},
$q^-$ poles appearing in $M$ are related to light-cone divergences that may be regulated
by choosing a slightly off-light like vector $n$.  Performing the contour integration 
under the described assumptions fixes the  momentum $q^{-}$ of the anti-quark
 in the loop in (\ref{eq:StartW})
to  $q^{-}=(\vec{q}_{T}^{2}+m_{s}^{2})/2q^{+}$.
The eikonal propagator can be split into a real and imaginary part 
via the usual principle value prescription $1/(x+i0)=\mathcal{P}(1/x)-i\pi\delta(x)$.
It has been argued\cite{Meissner:2007rx} that only the imaginary
part contributes to the relation (\ref{eq:Relation})
as it forces the antiquark momentum $q$ to be on the mass shell.
Thus, the imaginary part of the eikonal propagator corresponds
to a cut of the first diagram in Fig. \ref{fig:TMD-amplitude-including}
before the final-state interactions take place, whereas the real part
given by the principle value is  attributed
to terms that break the relation (\ref{eq:Relation}).   

After performing these integrations we  use (\ref{eq:StartW}) to calculate the pion Boer-Mulders function
via (\ref{eq:PhiSpectator}).  The pion-quark-antiquark vertex is
\begin{equation}
g_{\pi}(k^{2})=g_{\pi}\frac{(-\Lambda^{2})^{n-1}}{(n-1)!}\partial_{\Lambda^{2}}^{n-1}\frac{(k^{2}-m_{q}^{2})f(k^{2})}{k^{2}-\Lambda^{2}+i0} \  , 
\label{eq:Vertex}
\end{equation}
where the function $f$ is a homogeneous function of the quark virtuality. 
We choose it to be a Gaussian $\exp[-\lambda^{2}|k^{2}|]$\cite{Gamberg:2007wm}.
Inserting 
(\ref{eq:StartW})
into (\ref{eq:PhiSpectator}) 
yields the following expression for the Boer-Mulders function,
\begin{eqnarray}
\epsilon_{T}^{ij}k_{T}^{j}h_{1}^{\perp}(x,\vec{k}_{T}^{2})& = &\frac{2g_{\pi}^{2}m_{\pi}}{(2\pi)^{3}\Lambda^2}(xm_{s}+(1-x)m_{q})\left((1-x)\Lambda^{2}\right)^{2n-1}
\nonumber \\
 &  & \hspace{-3.0cm}\times\int\frac{d^{2}q_{T}}{(2\pi)^{2}}\frac{d^{2}p_{T}}{(2\pi)^{2}}
\epsilon_{T}^{ji}(q_{T}^{j}-p_{T}^{j})\frac{\mathrm{e}^{-\frac{2\lambda^{2}}{1-x}(xm_{s}^{2}-x(1-x)m_{\pi}^{2})}
\mathrm{e^{-\frac{\lambda^{2}}{1-x}(\vec{q}_{T}^{2}+\vec{p}_{T}^{2})}}}{\left[\vec{q}_{T}^{2}+\tilde{\Lambda}^{2}(x)\right]^{n}\left[\vec{p}_{T}^{2}+\tilde{\Lambda}^{2}(x)\right]^{n}}{\mathcal{F}}[{\bar{M}}^{\mathrm{eik}}] \  ,
\label{eq:BMNo}
\end{eqnarray}
where 
\begin{equation}
{\mathcal{F}}[{\bar{M}}^{\mathrm{eik}}]\equiv\left(\Im[\mathcal{\bar{M}}^{\mathrm{eik}}]\right)_{\delta\beta}^{\alpha\delta}(\vec{k}_{T}+\vec{q}_{T})\Big((2\pi)^{2}\delta^{\alpha\beta}\delta^{(2)}(\vec{p}_{T}+\vec{k}_{T})
+\left(\Re[\bar{{M}}^{\mathrm{eik}}]\right)_{\gamma\alpha}^{\beta\gamma}(\vec{k}_{T}+\vec{p}_{T})\Big)  
\end{equation}
with 
$
\tilde{\Lambda}^{2}(x)=xm_{s}^{2}-x(1-x)M^{2}+(1-x)\Lambda^{2}$.
This result already anticipates an eikonal form for the scattering amplitude
$\bar{{M}}(x,\vec{k}_{T},\vec{q}_{T})\rightarrow\bar{{M}}^{\mathrm{eik}}(|\vec{q}_{T}+\vec{k}_{T}|)
$
which we  exploit  to simplify the expression
and  show a relation to the chirally-odd GPD $H_{1}^{\pi}$. 
A calculation for the GPD $H_{1}^{\pi}$ 
for an antiquark spectator can be found in Ref.~\refcite{Meissner:2008ay},
which we generalize with the  vertex function $g_\pi(k^2)$
\begin{eqnarray}
\hspace{-.5cm}H_{1}^{\pi}(x,0,-\vec{\Delta}_{T}^{2}) &=&
 \frac{-g_{\pi}^{2}m_{\pi}}{2(2\pi)^{3}\Lambda^{2}}(xm_{s}+(1-x)m_{q})\left(\frac{(1-x)\Lambda^{2}}{\vec{D}_{T}^{2}+\tilde{\Lambda}^{2}(x)}\right)^{2n-1}
\nn &  & \hspace{-2cm}\times\int_{0}^{2\pi}d\varphi\int_{0}^{1}dz\,\frac{z^{2n-2}\mathrm{e}^{2\lambda^{2}\Lambda^{2}}\mathrm{e}^{-\frac{2\lambda^{2}(\vec{D}_{T}^{2}
+\tilde{\Lambda}^{2}(x))}{(1-x)z}}}{\left[1-4z(1-z)\frac{\vec{D}_{T}^{2}}{\vec{D}_{T}^{2}+
\tilde{\Lambda}^{2}(x)}\cos^{2}\varphi\right]^{n}},
\label{eq:H1Pi} 
\end{eqnarray}
where  $\vec{D}_{T}^{2}=\frac{1}{4}(1-x)^{2}\vec{\Delta}_{T}^{2}$.  Weighting (\ref{eq:BMNo})
with a transverse quark vector $k_{T}^{i}$ and integrating both sides
over $k_{T}$ we readily obtain the relation
\begin{equation}
m_{\pi}^{2}h_{1}^{\perp(1)}(x)=\int\frac{d^{2}q_{T}}{2(2\pi)^{2}}\,\vec{q}_{T}\cdot\vec{\mathcal{I}}(x,\vec{q}_{T})H_{1}^{\pi}\Big(x,0,-\left(\frac{\vec{q}_{T}}{1-x}\right)^{2}\Big)\,.
\label{eq:RelationDiquark}
\end{equation}
The function ${\mathcal{I}}^{i}$ can be expressed in terms of the real
and imaginary part of the scattering amplitude $\bar{{M}}$,
\begin{eqnarray}
\hspace{-0.5cm}{\mathcal{I}}^{i}(x,\vec{q}_{T})&=&
\frac{1}{N_{c}}\int\frac{d^{2}p_{T}}{(2\pi)^{2}}\,(2p_{T}-q_{T})^{i}\,
\left(\Im[\bar{{M}}^{\mathrm{eik}}]\right)_{\delta\beta}^{\alpha\delta}
(|\vec{p}_{T}|)
\nn &&
\hspace{-1cm}
\Big((2\pi)^{2}\delta^{\alpha\beta}\delta^{(2)}(\vec{p}_{T}-\vec{q}_{T})
+\left(\Re[\bar{{M}}^{\mathrm{eik}}]\Big)_{\gamma\alpha}^{\beta\gamma}(|\vec{p}_{T}-\vec{q}_{T}|)\right).
\label{eq:LensFunc}
\end{eqnarray}
In order to derive the relation (\ref{eq:Relation}) one transforms
Eq.~(\ref{eq:RelationDiquark}) into the impact parameter space
via a Fourier transform.
The lensing function in the impact parameter space then reads,
\begin{eqnarray}
\mathcal{I}^{i}(x,\vec{b}_{T}) &=&
 i(1-x)\int\frac{d^{2}q_{T}}{(2\pi)^{2}}\,\mathrm{e}^{i\frac{\vec{q}_{T}\cdot\vec{b}_{T}}{1-x}}I^{i}(x,\vec{q}_{T}).\label{eq:LensingFunctioneik}
\end{eqnarray}

\section{The Lensing  and Boer Mulders Function in Relativistic Eikonal Approximation}
In order to  
calculate the ${\bf 2\to 2}$
scattering amplitude $M$ (needed for (\ref{eq:LensFunc}))
we use functional methods to incorporate the color degrees of freedom
in the  eikonal limit when soft gauge bosons couple to highly
energetic  particles on the light cone. 
Here we summarize implementation of the color 
structure in the calculation of
 $M$ while the 
details of the functional approach can be found in a forthcoming publication.
Work in this
direction was carried out 
in Refs.~\refcite{Fried:1996uv,Fried:2000hj,Fried:2009fw}.  After some functional manipulations 
of the scattering amplitude,
$M$ can be expressed in terms of quark- and antiquark-propagators 
that are linked together by soft colored gluons. 
The amplitude ${M}$ reduces to a simple, gauge-invariant
expression\cite{Fried:2000hj}
\begin{eqnarray}
\left({M}^{\mathrm{eik}}\right)_{\delta\beta}^{\alpha\delta}(x,|\vec{q}_{T}+\vec{k}_{T}|) &=& \frac{(1-x)P^{+}}{m_{s}}\int d^{2}z_{T}\,\mathrm{e}^{-i\vec{z}_{T}\cdot(\vec{q}_{T}+\vec{k}_{T})}\label{eq:eikonalAmplitude}\\
 &  & \hspace{-3cm}\times\Bigg[\int d^{N_{c}^{2}-1}\alpha\int\frac{d^{N_{c}^{2}-1}u}{(2\pi)^{N_{c}^{2}-1}}\,\mathrm{e}^{-i\alpha\cdot u}\left(\mathrm{e}^{i\chi(|\vec{z}_{T}|)t\cdot\alpha}\right)_{\alpha\delta}\left(\mathrm{e}^{it\cdot u}\right)_{\delta\beta}-\delta_{\alpha\beta}\Bigg].\nonumber \end{eqnarray}
In Eq.~(\ref{eq:eikonalAmplitude}) the $N_{c}^{2}-1$  dimensional integrals over
the color parameters  results from  auxiliary fields $\alpha^{a}(s)$
and $u^{a}(s)$ that were introduced in the functional formalism of 
Ref.~\refcite{Fried:2000hj} in order to decouple the  gluon
fields from the color matrices.   The eikonal phase $\chi(|\vec{z}_{T}|)$ in Eq. (\ref{eq:eikonalAmplitude})
represents the arbitrary amount of soft gluon exchanges that are summed
up into an exponential form and is expressed in terms of the gluon
propagator in a covariant gauge,
\begin{equation}
\chi(|\vec{z}_{T}|)=g^{2}\int_{-\infty}^{\infty}d\alpha\int_{-\infty}^{\infty}d\beta\, n^{\mu}\bar{n}^{\nu}\mathcal{D}_{\mu\nu}(z+\alpha n-\beta\bar{n}).\label{eq:PhaseEik}\end{equation}
$\mathcal{D}$ denotes the gluon propagator, and $g$ the strong
coupling. In this form the four-vector $v$ is related to 
the complementary light cone vector $\bar{n}$, 
$v=-((1-x)P^{+}/m_{s})\bar{n}$,
with $n\cdot\bar{n}=1$ and $\bar{n}^{2}=0$. 
We  evaluate the color integrals by deriving a power series representation
for the color function
\begin{equation}
f_{\alpha\beta}(\chi)\equiv\int d^{N_{c}^{2}-1}\alpha\int\frac{d^{N_{c}^{2}-1}u}{(2\pi)^{N_{c}^{2}-1}}\,\mathrm{e}^{-i\alpha\cdot u}\left(\mathrm{e}^{i\chi(|\vec{z}_{T}|)t\cdot\alpha}\right)_{\alpha\delta}\left(\mathrm{e}^{it\cdot u}\right)_{\delta\beta}-\delta_{\alpha\beta}.\label{eq:ColorIntegral}
\end{equation}
After manipulating  the exponentials
in (\ref{eq:ColorIntegral}), 
rewriting the resulting factors as derivatives with respect to $u$, 
and performing an integration by parts 
we obtain the following power series representation for
$f$,
\begin{equation}
f_{\alpha\beta}(\chi)=\sum_{n=1}^{\infty}\frac{(i\chi)^{n}}{(n!)^{2}}\sum_{a_{1}=1}^{N_{c}^{2}-1}...\sum_{a_{n}=1}^{N_{c}^{2}-1}\sum_{P_{n}}\left(t^{a_{1}}...t^{a_{n}}t^{a_{P_{n}(1)}}...t^{a_{P_{n}(n)}}\right)_{\alpha\beta},
\label{eq:PowerSeries}
\end{equation}
where $P_{n}$ represents the sum over all permutations of the set $\{1,...,n\}$.
If we had  a direct ladder where gluons were not
allowed to cross we would 
have only factors $(t^{a_{1}}...t^{a_{n}}t^{a_{n}}...t^{a_{1}})_{\alpha\beta}=C_{F}^{n}\delta_{\alpha\beta}$
with $C_{F}=\frac{N_{c}^{2}-1}{2N_{c}}$, and we could work in an
Abelian theory with an effective replacement $\alpha\rightarrow C_{F}\alpha_{s}$
for the fine-structure constant. Since we allow generalized ladders
with crossed gluons we have to sum over all permutations in (\ref{eq:PowerSeries}),
and the simple  replacement is not possible. In a large $N_{c}$ expansion
the crossed gluons diagrams would be suppressed such that the direct
ladder represents the leading order in $1/N_{c}$.  In an Abelian theory, the generating 
matrices $t$ reduce to identity and since we have $n!$ permutations of the
set $\{1,...,n\}$, we  recover the well-known Abelian result,
\begin{equation}
f^{U(1)}(\chi)=\sum_{n=1}^{\infty}\frac{(i\chi)^{n}}{n!}=\mathrm{e}^{i\chi}-1.\label{eq:CFU1}
\end{equation}
For $N_{c}=2$, 
$t^{a}=\sigma^{a}/2$
and we can calculate the integral (\ref{eq:ColorIntegral}) analytically
by means of the relation $\left(\mathrm{e}^{iu\cdot\frac{\sigma}{2}}\right)_{\alpha\beta}=\delta_{\alpha\beta}\cos\left(\frac{|u|}{2}\right)+\frac{i\vec{\sigma}_{\alpha\beta}\cdot\vec{u}}{|u|}\sin\left(\frac{|u|}{2}\right)$.
We obtain,
\begin{equation}
f_{\alpha\beta}^{SU(2)}\left (\frac{\chi}{4} \right )=\delta_{\alpha\beta}\left(\cos\frac{\chi}{4}\,-\,\frac{\chi}{4}\sin\frac{\chi}{4}-1\right)+i\delta_{\alpha\beta}\left(2\sin\frac{\chi}{4}+\frac{\chi}{4}\cos\frac{\chi}{4}\right).\label{eq:SU2Analytical}\end{equation}
We also calculate numerically the lowest coefficients in the power
series (\ref{eq:PowerSeries}), and they exactly agree with the coefficients
in an expansion in $\chi$ of the analytical result (\ref{eq:SU2Analytical}).
This serves as a check of both numerical and analytical
approaches. For  $N_{c}=3$,  
due to  difficulty of
integrating over the Haar measure we use the power series (\ref{eq:PowerSeries}) to
 and obtain  the following approximative color function which is valid
if $a=\chi/4$ is small,
\begin{eqnarray}
\Re[f_{\alpha\beta}^{SU(3)}](a)&=&
 \delta_{\alpha\beta}(-c_{2}a^{2}+c_{4}a^{4}-c_{6}a^{6}-c_{8}a^{8}+...),
\nn
\Im[f_{\alpha\beta}^{SU(3)}](a) &=& \delta_{\alpha\beta}(c_{1}a-c_{3}a^{3}+c_{5}a^{5}-c_{7}a^{7}+...),\label{eq:CFSU3Im}
\end{eqnarray}
with the numerical values 
$c_{1}=5.333$, $c_{2}=6.222$, $c_{3}=3.951$,
$c_{4}=1.934$, $c_{5}=0.680$, $c_{6}=0.198$, $c_{7}=0.047$, $c_{8}=0.00967$.%
\begin{figure}
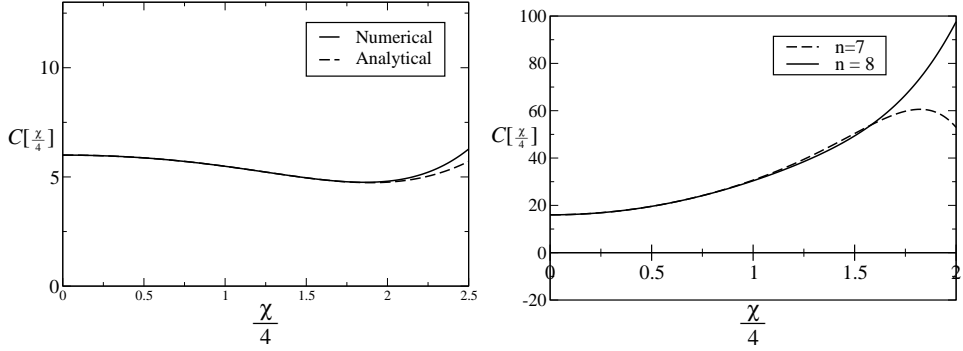

\begin{center}
\vskip .25cm
\includegraphics[scale=0.25]{SU2Color}~~\includegraphics[scale=0.25]{SU3Color}
\caption{\scriptsize The function $C[\frac{\chi}{4}]$ of Eq. (\ref{eq:Cchi}) as a function
of the eikonal amplitude $\frac{\chi}{4}$. Left: $SU(2)$, we 
compare the numerical result computed by means of Eq. (\ref{eq:PowerSeries})
up to the order $n=8$ with the analytical result in Eq. (\ref{eq:LensIPSU2}).
The numerical agrees with the analytical result up to $\frac{\chi}{4}\sim2$.
Right: $SU(3)$, we compare the numerical results for the
orders $n\,=7,\,8$. The results are accurate
up to $\frac{\chi}{4}\sim1.5$.\label{fig:The-function-} }
\end{center}
\end{figure}
Working in coordinate space where  we  can express the lensing 
function directly in terms of the real and imaginary part of the
color function $f$ which is itself a function of the eikonal
phase $\chi$ Eq. (\ref{eq:PhaseEik}) , results in 
 a lensing function of the form,
\begin{equation}
  {\mathcal{I}}^{i}(x,\vec{b}_{T})=
\frac{(1-x)}{2N_{c}} \frac{b_{T}^{i}}{|\vec{b}_{T}|}
\frac{\chi^{\prime}}{4} C \left [\frac{\chi}{4} \right ] \  ,\label{eq:LensColor} 
\end{equation}
with  
\begin{align}
 C\left [\frac{\chi}{4} \right ]\equiv 
&
 \Bigl [\left(\Tr\Im[f]\right)^{\prime}\left (\frac{\chi}{4} \right )+ 
\frac{1}{2}\Tr\left[\left(\Im[f]\right)^{\prime}\left (\frac{\chi}{4} \right )
\left(\Re[f]\right)\left (\frac{\chi}{4} \right )\right]
\nonumber \\
&
-\frac{1}{2}\Tr\left[\left(\Im[f]\right)\left (\frac{\chi}{4} \right )
\left(\Re[f]\right)^{\prime}\left (\frac{\chi}{4} \right )\right]\Bigr]
 \ ,\label{eq:Cchi}
\end{align}
where $\chi^{\prime}$ denotes the first derivative with respect to
$|\vec{z}_{T}|$, and $\left(\Im[f]\right)^{\prime}$ and $\left(\Re[f]\right)^{\prime}$
are the first derivatives of the real and imaginary parts of the color
function $f$. 
Inserting (\ref{eq:CFU1}) into (\ref{eq:LensColor})
yields the lensing function in
an Abelian $U(1)$-theory 
\begin{equation}
\mathcal{I}_{U(1)}^{i}(x,\vec{b}_{T})=(1-x)\frac{b_{T}^{i}}{4|\vec{b}_{T}|}\chi^{\prime}(\frac{|\vec{b}_{T}|}{1-x})\left(1+\cos\chi(\frac{|\vec{b}_{T}|}{1-x})\right).\label{eq:LensIPU1}\end{equation}
Likewise by using (\ref{eq:SU2Analytical})  the lensing
function in an $SU(2)$-theory is given by
\begin{equation}
\mathcal{I}_{SU(2)}^{i}(x,\vec{b}_{T})=\frac{(1-x)b_{T}^{i}}{16|\vec{b}_{T}|}
\chi^{\prime}
\left(3(1+\cos\frac{\chi}{4})+\left(\frac{\chi}{4}\right)^{2}-\sin\frac{\chi}{4}\left(\frac{\chi}{4}-\sin\frac{\chi}{4}\right)\right),\label{eq:LensIPSU2}\end{equation}
where $\chi=\chi\left(\frac{|\vec{b}_{T}|}{1-x}\right)$.
In an $SU(3)$-theory, we use 
the approximate color function $f$ 
to calculate the lensing function as a function
of the eikonal amplitude.
In Fig. \ref{fig:The-function-}
the function $C[\frac{\chi}{4}]$ is plotted versus $\frac{\chi}{4}$
for various approximations. While the convergence of the power series
seems to be better for $SU(2)$ than in the $SU(3)$ case where the
numerical result calculated with eight coefficients agrees with the
analytical result up to $\frac{\chi}{4}\sim2$, we can trust
 the numerical result computed with eight coefficients up
to $\frac{\chi}{4}\sim1.5$ for $SU(3)$.

\begin{figure}
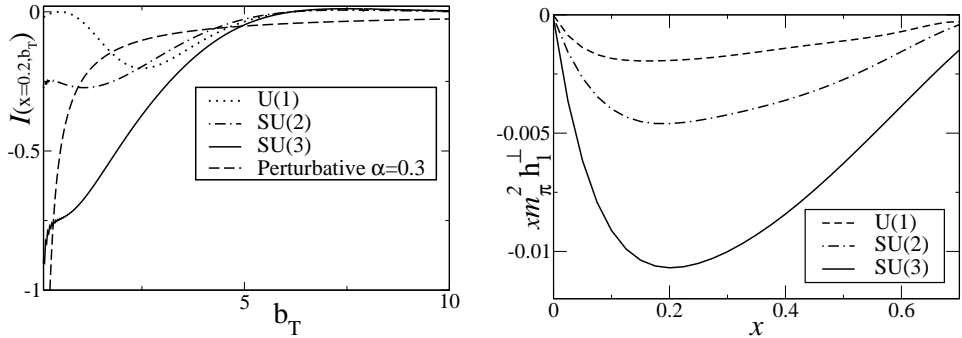

\begin{center}
\vskip 0.25cm
\includegraphics[scale=0.25]{LensingIP}~~~~\includegraphics[scale=0.25]{PionBM}
\caption{Left:
The lensing function $\mathcal{I}^{i}(x,\vec{b}_{T})$
from Eq. (\ref{eq:LensColor}) for $U(1)$, $SU(2)$ and $SU(3)$
for $x=0.2$ at a scale $\Lambda_{QCD}=0.2\,\mathrm{GeV}$. For comparison
we also plot the perturbative result with an
arbitrary value for the coupling, $\alpha=0.3$.
Right: First moments of the pion Boer-Mulders function calculated by means
of the relation to the chirally-odd GPD $\mathcal{H}_{1}^{\pi}$ for
a $SU(3)$, $SU(2)$, and  $U(1)$ gauge theory.
\label{fig:The-eikonal-phase}}
\end{center}
\end{figure}

In order to numerically estimate the lensing function
and in turn the Boer Mulders function 
we seek to utilize the infrared behavior of the gluon
and the running coupling in the non-perturbative regime where we
  infer
that the soft gluon transverse momentum defines the scale at which
the coupling is evaluated. 
These two quantities 
have been extensively studied in the infrared limit in the
 Dyson-Schwinger framework\cite{Alkofer:2008tt}
and in lattice QCD\cite{Sternbeck:2008mv}. 
We use calculations of these quantities from  Dyson-Schwinger equations\cite{Alkofer:2008tt}
where both $\alpha_s$ and $\mathcal{D}^{-1}$ 
are defined in the infrared limit (details can be found in a forthcoming publication).
This determines the eikonal phase and thus the lensing
functions (\ref{eq:LensColor}) for a $U(1)$, $SU(2)$ and $SU(3)$
color function. We plot the results in Fig. \ref{fig:The-eikonal-phase}
for a color function for $U(1)$, $SU(2)$, $SU(3)$. While we observe
that all lensing functions are attractive and 
fall off at large transverse distances,
they are very different in size at small distances.

Using  the eikonal model for the lensing
function together with the  spectator model for the GPD $H_{1}^{\pi}$
we present predictions of the relation (\ref{eq:Relation}) 
for the first moment of the pion Boer-Mulders
function $h_{1}^{\perp(1)}$.
We  fix the 
six free model parameters $m_{s}$, $m_{q}$, $\Lambda$,
$\lambda$, $g_{\pi}$ and $n$ in (\ref{eq:H1Pi}) 
that we need to determine by fitting
to pion data .
In order to do so we determine the chiral-even GPD $F_{1}^{\pi}$
in the spectator model by
 investigating  different limits of $F_{1}^{\pi}$. When integrated
over $x$, the GPD reduces to the pion form factor $F^{\pi^{+}}(Q^{2})=-F^{\pi^{-}}(Q^{2})$.
An experimental fit of the Pion form factor to data is presented in
Refs.~\refcite{Blok:2008jy,Huber:2008id}, and up to $Q^{2}=2.45\,\mathrm{GeV^{2}}$
a reasonable fit to the data is displayed by the monopole formula
$F_{\mathrm{fit}}(Q^{2})=(1+1.85\, Q^{2})^{-1}$. This procedure is
expected to predict the $t$-dependence of the chirally-odd GPD $H_{1}^{\pi}$
reasonably well up to $Q^{2}=2.45\,\mathrm{GeV^{2}}$. In order to
fix the $x$-dependence of $H_{1}^{\pi}$ we fit the collinear limit
$F_{1}^{\pi}(x,0,0)$ to the valence quark distribution in a pion,
$v(x)$. A parameterization for this object was given for example
by GRV in Ref.~\refcite{Gluck:1991ey} at a scale $\mu^{2}=2\,\mathrm{GeV^{2}}$.
Reasonable agreements of the Form Factor- 
$F_{1}^{\pi}(x,0,0)$ 
with the data fits were obtained for the parameters $m_{q}=0.834\,\mathrm{GeV}$,
$m_{s}=0.632\,\mathrm{GeV}$, $\Lambda=0.067\,\mathrm{GeV}$, $\lambda=0.448\,\mathrm{GeV}$,
$n=0.971$, $g_{\pi}=3.604$.  With the predicted GPD $H_{1}^{\pi}$ and the lensing function 
$\mathcal{I}^{i}(x,\vec{b}_{T})$ 
as input we can use the relation (\ref{eq:Relation}) to give a prediction
for the valence contribution to the first $k_{T}$-moment of the pion
Boer-Mulders function, which we can write as,
\begin{equation}
m_{\pi}^{2}h_{1}^{\perp(1)}(x)=2\pi\int_{0}^{\infty}db_{T}\, b_{T}^{2}\mathcal{I}(x,b_{T})\frac{\partial}{\partial b_{T}^{2}}\mathcal{H}_{1}^{\pi}(x,b_{T}^{2}).
\end{equation}
We present numerical results for $xm_{\pi}^{2}h_{1}^{\perp(1)}(x)$
shown in Fig. \ref{fig:The-eikonal-phase} for a $U(1)$, $SU(2)$ and $SU(3)$
gauge theory~\cite{Gamberg:2009uk}.   
It was argued in Ref.~\refcite{Burkardt:2003uw}
that a negative sign of the lensing functions indicates attractive
FSIs. We find that the lensing function is negative for the
both the Abelian and  non-Abelian gauge theories. 
The magnitude of the $SU(3)$ result
is about $0.01$, while the $SU(2)$ result and $U(1)$ result are
smaller. One observes a growth
of the pion Boer-Mulders function with $N_{c}$ which was also predicted
by a model-independent large $N_{c}$ analysis\cite{Pobylitsa:2003ty}.
So far the pion Boer-Mulders function is  unknown but maybe determined 
 from a future pion-proton Drell-Yan experiment to be
performed at COMPASS. Once a pion Boer-Mulders is extracted
our analysis can be used to verify quantitatively
GPD - TMD relations. An extraction of the other T-odd parton distribution,
the proton Sivers function $f_{1T}^{\perp(1)}$, from SIDIS data measured
at HERMES and COMPASS reveals an effect of the magnitude of about
$0.04$, four times larger than our prediction. 
 A similar calculation for the
proton Sivers function will be reported elsewhere.
 
\section*{Acknowledgments} 
L.G.  dedicates this paper to  Anatoli  Efremov on the occasion of his 75th 
birthday. I am  grateful to the organizers
of the workshop���� Recent Advances in 
Perturbative QCD and Hadronic Physics ECT$^{\star}$,
Trento (Italy) for their efforts which made this  memorable event possible. 
L.G. acknowledges support from  U.S. Department of Energy under 
contract DE-FG02-07ER41460.  Notice: Authored by Jefferson Science 
Associates, LLC under U.S. DOE Contract No. 
DE-AC05-06OR23177. The U.S. Government retains a non-exclusive, paid-up, 
irrevocable, world-wide license to publish or reproduce this manuscript for U.S. 
Government purposes.

\bibliographystyle{slac} 
\bibliography{Referenzen}
 
\end{document}